\documentclass{ws-ijgmmp}
\usepackage{amsfonts,amssymb,latexsym}

\usepackage{color}
\usepackage{colordvi}
\usepackage{amssymb,amsmath}
\usepackage{hyperref}
\usepackage{graphics}
\usepackage{epsfig}
\newcommand{\beq}{\begin{equation}}
\newcommand{\eeq}{\end{equation}}
\newcommand{\cc}{{\rm c}}

\newcommand{\diff}{\text{d}\,}
\newcommand{\display}{\displaystyle}

\def\bar#1{\begin{array}{#1}}
\def\ear{\end{array} }

\def\display{\displaystyle}

\def\foot{\footnote}

\def\pmb#1{\setbox0=\hbox{$#1$}%
 \kern-.025em\copy0\kern-\wd0
 \kern.05em\copy0\kern-\wd0
 \kern-.025em\raise.0433em\box0}

\def\centerwmf#1#2#3{\vskip#2\relax\centerline{\hbox to#1{\special {wmf:#3 x=#1, y=#2}\hfil}}}

\def\bar{\overline}
\def\text{\rm}
\def\text{\mbox}
\begin{document}

\markboth{I. Bochicchio, S. Capozziello and E. Laserra} {The
Weierstrass criterion and the LTB Models}

%
\catchline{}{}{}{}{}
%

\title{THE WEIERSTRASS CRITERION AND THE
LEMA$\hat{\text{I}}$TRE--TOLMAN--BONDI MODELS WITH COSMOLOGICAL
CONSTANT $\Lambda$}

\author{IVANA BOCHICCHIO}

\address{Dipartimento di Matematica ed Informatica, Universit\'a
degli Studi di Salerno\\
Via Ponte Don Melillo, 84084, Fisciano (SA), Italy\\
\email{ibochicchio@unisa.it} }

\author{SALVATORE CAPOZZIELLO}

\address{Dipartimento di Scienze Fisiche, Universit\'a
di Napoli`` Federico II'' and INFN Sez. di Napoli, \\
Compl. Univ. di Monte S. Angelo, Edificio G,
Via Cinthia, I-80126, Napoli, Italy\\
capozzie@na.infn.it }

\author{ETTORE LASERRA}

\address{Dipartimento di Matematica ed Informatica, Universit\'a
degli Studi di Salerno\\
Via Ponte Don Melillo, 84084, Fisciano (SA), Italy\\
\email{elaserra@unisa.it} }

\maketitle


\begin{abstract}
We analyze Lema{\^{\i}}tre--Tolman--Bondi models in presence of
the cosmological constant $\Lambda$ through the classical
Weierstrass criterion. Precisely, we show that the Weierstrass
approach allows us to classify the dynamics of these inhomogeneous
spherically symmetric Universes taking into account their
relationship with the sign of $\Lambda$.
\end{abstract}


\section{Introduction}
The Weierstrass criterion is one of the most elegant and powerful
tools of classical mechanics, in that it allows qualitative
analysis of one-dimensional conservative motions, once the zeros
of a suitable function, the Weierstrass function, are determined.
The usefulness of this criterion also lies in the fact that it can
also be successfully generalized to many problems outside of
classical mechanics, see e.g. the recent papers
\cite{jim,grg,saccomandi}.
\par
In particular, the focus of this paper is to study the qualitative
behavior of Lema{\^{\i}}tre-Tolman-Bondi (LTB) models, endowed
with a non-null cosmological constant $\Lambda$, by assuming a
Weierstrass approach. \footnote{\ A similar approach has been
developed in \cite{jim,grg}, but in that case only LTB models with
null $\Lambda$ were considered.}
\newline
The importance of such Universes in relativistic cosmology is
mainly based on the following observation:
even if the Friedmann-Robertson-Walker (FRW) models\,%
\footnote{\ That is the  homogeneous spherically symmetric dust
models.} are widely accepted to describe the cosmic evolution,
there are several objections to such models which include
singularities, horizons, observed inhomogeneities starting from
galactic scales up to galaxy superclusters (e.g. Virgo
supercluster) \cite{RAO-ANNAPURNA}.
%
%
In order to overcome some of the difficulties faced by the FRW
models, Lema{\^{\i}}tre \cite{Lemaitre933}, Tolman \cite{tolman},
Bondi \cite{bondi947} and others have considered inhomogeneous
spherically symmetric dust models. Hence, the main aim of the
LTB approach is to encompass cosmic inhomogeneities, at both large
and small scales, with the overall cosmic dynamics
\cite{celerier}. Besides, being the simplest inhomogeneous
solutions of the Einstein equations, it is relatively easy to
work with them. Furthermore, the interest in such models is
recently increased due to the fact that some of them can be
designed to satisfy several observational requirements
\cite{krasinski-hellaby1,krasinski-hellaby2,krasinski-hellaby3}.

As such, through the Weierstrass method, we are able to
qualitatively describe and classify the possible kinds of
evolution of the LTB--models. In particular, in the peculiar case
of the FRW--models, Friedmann discussion \cite{friedman,plebanski}
can be reduced to a straightforward application of the Weierstrass
method.
\par
\par
The paper is organized as follows: in Sec. 2, the main features of
LTB Universes are reviewed.
A discussion on the qualitative study of the evolution of the
$r$--shells in General Relativity through the Weierstrass method
is given in Sec. 3. Section 4 is devoted to concluding remarks.

\section{The Lema{\^{\i}}tre--Tolmann--Bondi Models}
Let us give now a brief summary of the main features of LTB-models
according to \cite{rassegna,bondi947,laserra982,laserra985}. We
will consider a dust system $C$
which, during its evolution, generates a Riemannian manifold,
with locally spatial spherical symmetry around a physical point $O$;\,%
\footnote{\ See \cite{levi-civita926} for a precise definition of
locally spatial spherical symmetry around a physical point $O$.}
the metric can then be given the Levi Civita's form
\cite{levi-civita926}:
\begin{equation}
\diff s^{2}=A^{2}(t,r)\diff r^{2}+R^{2}(t,r)(\diff
\theta^{2}+\sin^{2}\theta d\varphi^{2})-\cc^2\,\diff t^{2}\ ,
\end{equation}
where $t$ is the proper time of each particle and $r$, $\theta$,
$\varphi$ are co--moving Levi--Civita's curvature spherical
coordinates: we can interpret $R(t,r)$ as the intrinsic radius of
the $O$--sphere $\mathcal{S}(r)$ at time $t$ so that
${\displaystyle \frac{1}{R^2}}$ represents, at any point, the
Gaussian curvature of the geodesic sphere with its centre at $O$.

We consider now the initial space-like hypersurface $V_3$ (with
equation $t=0$) and call $r$--shells the set of particles with
co--moving radius $r$ (i.e. the dust initially distributed on the
surface of the geodesic sphere with center at $O$ and radius $r$
($O$--sphere) $\mathcal{S}(r)$). According to
\cite{rassegna,laserra982,laserra985}, we assign, at each particle
of a $r$--shell, the initial intrinsic radius $r=R(r,0)$ as the
radial co--moving coordinate. Hence the metric of the initial
spatial manifold $V_3$ takes the form
\begin{equation}
\diff \sigma^{2}=a^{2}(r)\diff r^{2}+r^{2}(\diff
\theta^{2}+\sin^{2}\theta\diff\varphi^{2})\ ,
\end{equation}
where $a(r)=A(r,0)$.
\par
Now let us take into account the gravitational field equations
with the cosmological constant $\Lambda$ and the conservation
equations:
\begin{equation}\label{eq:field-lambda}
\left\{\begin{split} G_{\alpha\beta}+\Lambda\,g_{\alpha\beta} &= -
\display\frac{8\,\pi\,G_N}{\cc^4} T_{\alpha\beta}\,,
\\[1ex]
\nabla_{\alpha}T^{\alpha}_{\beta} &= 0\,,
\end{split}\right.\ ,
\end{equation}
where $G_N$ is the Newton gravitational constant.
\newline
It has been shown \cite{caricato968} that it is possible to break
the corresponding Cauchy problem into two separate intrinsically
formulated invariant problems: the problem of initial conditions
and the restricted problem of evolution. By taking into account
\cite{rassegna,bondi947,caricato968,laserra982,laserra985}, we can
immediately translate the restricted evolution problem
into the following equations\,%
\foot{\ These equations differ from the previous
\cite{laserra982,laserra985} only for the additional term
$\frac{\Lambda}{3}\,{R}^{2}$.}
\begin{equation}\label{(1.5)-Lambda}
\left\{
\begin{array}{lll}
A(t,r) & = & a(r)\,R'(t,r)
\\[1ex]
\dot{R}^{2} & = & \display\cc^2\,\left(\frac{1}{a^{2}(r)}-1\right)
+ \frac{2\,G_N\,M(r)}{R} + \frac{\Lambda}{3}\,{R}^{2}
\\[1.5ex]
\mu(t,r) & = & \display\frac{\mu_{0}(r)r^{2}}{R'(t,r)R^{2}(t,r)}
\end{array}
\right.
\end{equation}
where a dot denotes differentiation with respect to $t$ and a
prime differentiation with respect to $r$, $\mu(t,r)$ is the mass
density ($\mu_{0}=\mu(r,0)$)
and $M(r)$ is the ``\textit{Euclidean mass}''\,%
\foot{\ In that $M(r)$ would be the mass of dust contained within
$\mathcal{S}(r)$, if the initial $O$--sphere $V_3$ was Euclidean.}
\begin{equation}\label{(1.3)}
M(r)=4\pi \int_{0}^{r}\mu_{0}(s)\,s^{2}\,\diff s\ .
\end{equation}
\indent
If the initial mass density is constant, equation
\eqref{(1.5)-Lambda}$_2$ reduces to the Friedman equation and
$M(r)$ is constant.
\par
If, analogously to \cite{bo-la006}, we introduce the function
\begin{equation} \label{epsilonDefinition}
\varepsilon(r)=\frac{1}{a^2(r)}-1\ ,
\end{equation}
which represents the percentage deviation of $a^2(r)$ from the
Euclidean value $a^2=1$, Eq. \eqref{(1.5)-Lambda}$_2$ becomes
\begin{equation}\label{eq:3.3'ex4.3'}
\dot{R}^{2} = \display\varepsilon(r)\,\cc^2+\frac{2\,G_N\,M(r)}{R}
+ \frac{\Lambda}{3}R^{2}\,.
\end{equation}
\begin{remark}
The function $\varepsilon(r)$ and the initial metric coefficient
$a^2(r)$ are connected to the initial data by the equation
\begin{equation}\label{(1.6)-Lambda}
\varepsilon(r) = \frac{1}{\cc^2}\,\left(\dot{R}_0^2(r) -
\frac{2\,G_N\,M(r)}{r} - \frac{\Lambda}{3}\,r^{2}\right)\,
\end{equation}
which gives an initial constraint.
\end{remark}

\section{Generalization of the Weierstrass Criterion to the LTB--Models}

\subsection{The Weierstrass equation}
We will use a generalized form of the Weierstrass criterion,\,%
\foot{\ For a detailed exposition of the Weierstrass criterion see
e.g. \cite[\S\,1.3 p.23]{benenti}.} which can be applied to many
problems outside of classical mechanics too. Let's consider the
first order differential equation \beq\label{eq:Weierstrass}
\dot{x}^{2} = \Phi({x})\ , \eeq that we may call
\textit{Weierstrass equation} with \textit{Weierstrass function}
$\Phi({x})$ (see \cite{benenti}).
Equations of the form \eqref{eq:Weierstrass} are frequently
encountered in classical mechanics. For example, the natural
motions of a material point of mass $m$, subjected to a
conservative force deriving from the potential energy $V(x)$ and
having mechanical energy $E$, are described by the Weierstrass
equation \beq\label{eq:Weierstrass'} \dot{x}^{2} =
\frac{2}{m}[E-V(x)]\ , \eeq whith the Weierstrass function
\[
\Phi({x}) = \frac{2}{m}[E-V(x)]\ .
\]

In addition the Weierstrass Eq. \eqref{eq:Weierstrass} translates
into the double equation
\begin{equation}\label{eq:sqrtWeierstrass}
\frac{\diff {x}}{\diff t} = \pm\sqrt{\Phi({x})}\ ,
\end{equation}
which can be integrated by separating the variables
\begin{equation} \label{eq:intGenWeie}
t(x) = \pm\int_{x_0}^x\frac{\diff x}{\sqrt{\Phi(x)}} + t_0
\end{equation}
where we choose the sign $\pm$ in agreement with the sign of the
initial rate $\dot{x}_0$,
\[
\dot{x}_0^{2} = \Phi(x_0)\ .
\]
\indent
The importance of the Weierstrass approach is mainly based on the
fact that it is possible to
%
%
obtain the qualitative behavior of the solutions of a Weierstrass
equation, without integrating it. Precisely, the zeros of the
Weierstrass function have a leading role,
in fact the solutions of the Weierstrass equation are confined in
those regions of the ${x}$--axis where $\dot{x}^2\geq 0$, hence
the Weierstrass condition
\begin{equation}\label{eq:PhiNonnegative}
\Phi({x}) \geq 0
\end{equation}
must be fulfilled: \textit{the solutions of the Weierstrass
equation are confined in those regions where the Weierstrass
function is non negative}.
\newline
These regions are unlimited or are limited by the extrema of the
definition interval (eventually $+\infty$ or $-\infty$) and by the
eventual zeros of the Weierstrass function, which are called
\textit{barriers}, because they cannot be crossed by the solution
$x(t)$, so they split the range of possible values for $x$ into
\textit{allowed} and \textit{prohibited} intervals.
\newline
Now let's consider the zeros of $\Phi$.
\par
If a barrier $x_B$ is a simple zero of $\Phi$, that is such that
\[
\Phi({x}_{B}) = 0\ ,\quad\Phi'({x}_{B})\neq 0\ ,
\]
then it is called an \textit{inversion point} ${x}_{I}$, because
the motion reverses its course after reaching it
(see e.g. \cite{benenti,saccomandi}).\,%
\par
If a barrier ${x}_{B}$ is a multiple zero, that is such that
\[
\Phi({x}_{B})=0\ ,\quad\Phi'({x}_{B}) = 0\ ,
\]
then it separates two allowed intervals and it is called a
\textit{soft barrier} ${x}_{S}$ \cite{saccomandi}.
\newline
A soft barrier is also called an \textit{asymptotic point},
because it takes an infinite time to reach it. In fact at a soft
barrier the integral \eqref{eq:intGenWeie} diverges (see e.g.
\cite{benenti,saccomandi}).
\newline
Finally we recall that \textit{an asymptotic point is also an
equilibrium point} (see e.g. \cite{benenti}).
\par
So, once these zeros are found, the qualitative behavior of the
solutions of the Weierstrass equation \eqref{eq:Weierstrass} is
completely determined.

\subsection{The Weierstrass criterion for the evolving $r$--shells}
In order to qualitatively study the behavior of $r$--shells in the
case of non--null cosmological constant, let us analyze Eq.
\eqref{(1.5)-Lambda}$_2$ which determines the evolution of the
material continuum through the Weierstrass method.\,%
\foot{\ Once solved Eq. \eqref{(1.5)-Lambda}$_2$, Eqs.
\eqref{(1.5)-Lambda}$_1$,\eqref{(1.5)-Lambda}$_3$ can be
immediately solved.}
\newline
Now we will focus our attention on a given single $r$-shell (that
is we will consider $r$ as a fixed parameter), so we can regard
the intrinsic radius $R$ as a function of the time only,
$x(t)=R(t;r)$, and $\varepsilon(r)\,,M(r)$ as constant; then Eq.
\eqref{eq:3.3'ex4.3'} becomes a quadratic differential equation
\begin{equation} \label{weie2-ripetuto}
\dot{x}^{2}=\varepsilon\,\cc^2+\frac{2\,G_N\,M}{x}+\frac{\Lambda}{3}\,{x}^{2}
\ ,
\end{equation}
that is a Weierstrass equation with Weierstrass function
\beq\label{eq:WeierstrassPhi} \Phi({x};r) =
\varepsilon(r)\,\cc^2+\frac{2\,G_N\,M(r)}{x}+\frac{\Lambda}{3}\,{x}^{2}
\ , \eeq depending on the parameter $r$. Eq.
\eqref{weie2-ripetuto} translates into two equations, depending on
the parameter $r$,
\begin{equation}\label{eq:sqrtWeierstrass}
\frac{\diff {x}}{\diff t} = \pm\sqrt{\Phi({x};r)} =
\pm\sqrt{\varepsilon(r)\,\cc^2+\frac{2\,G_N\,M(r)}{x}+\frac{\Lambda}{3}\,{x}^{2}}
\ ,
\end{equation}
where we have to choose the sign $\pm$ in agreement with the sign
of the initial rate $\dot{x}_0$,
\[
\dot{x}_0^{2} = \Phi(x_0;r) = \varepsilon(r)\,\cc^2 +
\frac{2\,G_N\,M(r)}{r} + \frac{\Lambda}{3}\,r^2 \ .
\]
\begin{remark}
In consequence of Poincar\'e's theorem on the analytic dependence
of a solution on a parameter, the solutions of each of the Eqs.
\eqref{eq:sqrtWeierstrass} depend analytically on the parameter
$r$ in each point where $\sqrt{\Phi}$ depends analytically on
${x}$ and $r$ \rm{(see e.g. \cite{elsgolts})}.
\end{remark}
\indent The zeros of the Weierstrass function
\eqref{eq:WeierstrassPhi} depend not only on the sign of
$\Lambda$, but obviously on the sign of $\varepsilon$ too, so the
evolution of the $r$--shells will depend on both signs, the sign
of $\varepsilon$ and the sign of $\Lambda$. These different
situations will be analyzed in details in the following sections,
where the zeros of the Weierstrass function are obtained by
finding the positive real roots of the third degree equation
\begin{equation}\label{terzo_grado}
\Lambda x^{3}+3\varepsilon c^{2}x+6G_{N}M=0\ .
\end{equation}
To highlight the role of the cosmological constant, in the
following we will consider Eq. \eqref{weie2-ripetuto} in the form:
\begin{equation} \label{Phi_nuovo}
\dot{x}^2
= \frac{1}{3}\,x^2\left[\Lambda+W(x)\right]\ ,
\end{equation}
where
\begin{equation} \label{W}
W(x)\,=\,\frac{3 \varepsilon \cc^2}{x^2}+\frac{6 G_N M}{x^3}\ .
\end{equation}
Then the Weierstrass condition \eqref{eq:PhiNonnegative} holds
when
\begin{equation}\label{segnoW}
\Lambda \geq - W(x).
\end{equation}
\begin{remark}\label{osservazione}
Once $W(x)$ is introduced, a barrier $x_{B}\,(\neq 0)$
is a soft barrier $x_{S}$ iff $\,W^{\prime}(x_B)\,=\,0$.
\end{remark}
In fact from Eqs.\eqref{eq:Weierstrass} and \eqref{Phi_nuovo} it
follows that
$$\Phi^{\prime }(x)=\frac{2}{3}x\left[ \Lambda +W(x)\right]
+\frac{1}{3}x^{2}W^{\prime }(x).$$ Supposing $\Phi(x_B)=0$ then
$\Lambda+W(x_B)=0$; so, since
$\Phi^{\prime}(x_B)=\frac{1}{3}x^{2}W^{\prime }(x_B)$,
$\Phi^{\prime}(x_B) = 0 \Rightarrow W^{\prime}(x_B) = 0$ and
viceversa.
\par
We remark that, by considering the Weierstrass function $\Phi(x)$
in the peculiar case of a FRW--Universe, it is easy to re--obtain
the Friedmann discussion \cite{friedman}.

\subsubsection{Null cosmological constant $\Lambda$}

For sake of completeness, in this section we briefly recall the
qualitative behavior of LTB--models with $\Lambda\,=\,0$, since
they have already been studied in \cite{jim,grg}.
\par
For negative values of $\varepsilon$, there is only one barrier
\[
x_I = -\frac{2GM}{\varepsilon\cc^2}\ ,
\]
which is a simple zero of the Weierstrass function, that is an
inversion point. So, if a $r$--shell is initially expanding, it
will go on expanding until the intrinsic radius reaches the value
$x_I$; then it will contract back from $x_I$ towards the center of
symmetry $O$ until it collapses in a finite time (see \cite{jim}).
\par
When $\varepsilon$ is null (Euclidean case) the inversion point
goes to infinity, so if a $r$--shell is initially expanding, it
will go on expanding without limit, approaching the null expansion
rate (see \cite{jim}).
\par
Finally, a similar result can be achieved for positive
$\varepsilon$: there are no barriers, so if the $r$--shell is
initially expanding it will go on expanding with decreasing rate
approaching the limit value
$\dot{x}_l\,=\,\sqrt{\varepsilon\cc^2}$ (see \cite{jim}).

\begin{figure}[h]
\centering
\includegraphics[width=0.87\textwidth]{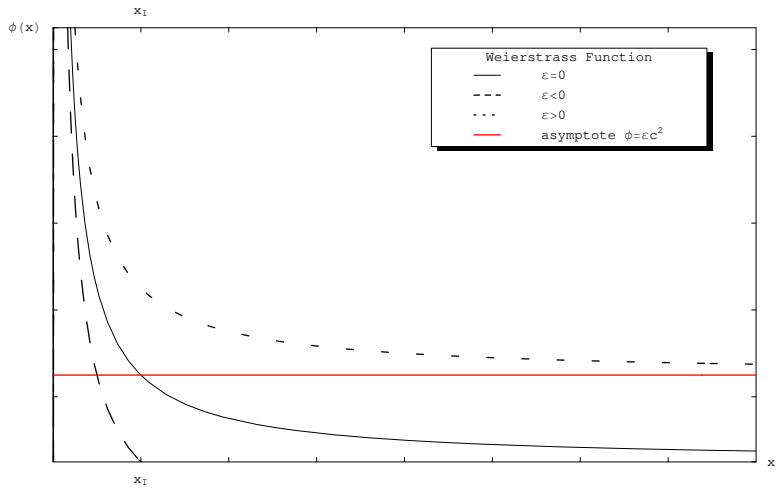}
\caption{ The Weierstrass function $\Phi({x})$ for $\Lambda=0$ in
three different cases. More precisely
\newline
i) when $\varepsilon < 0$, the function $\Phi({x})$ has a simple
zero at ${x}_{_I}$ (i.e. there is one inversion point ${x}_{_I}$)
and the motion $\forall \,\, 0 < {x} \leq {x}_{_I}$ is possible;
\newline
ii) when $\varepsilon = 0$, the inversion point goes to infinity
and the motion is possible $\forall\; x>0$;
\newline
iii) when $\varepsilon > 0$, there are no barriers: the motion
$\forall\, x > 0$ is possible. In this case the Weierstrass
function admits the red line $\Phi(x)\,=\,\varepsilon \cc^2$ as
asymptote.}
\end{figure}

\subsubsection{Positive cosmological constant $\Lambda$}

The positivity of the cosmological constant implies an open model
for null and positive values $\varepsilon(r)$, while a different
and more complex situation is obtained for negative values of
$\varepsilon(r)$.

\subsubsection*{$\Lambda$ is positive and $\varepsilon$ is positive or null}
The evolution of LTB--models in these two cases is very similar.
\par
In fact when $\varepsilon\,=\,0$ (Euclidean case), the function
\eqref{W} becomes
\begin{equation} \nonumber
W(x)\,=\,\frac{6 G_N M}{x^3} \ ;
\end{equation}
so, from \eqref{segnoW}, if $\Lambda>0$ then $\Phi({x}) >
0\;\forall x>0$.
\par
Instead when $\varepsilon\,>\,0$, the function \eqref{W} becomes
\begin{equation} \nonumber
W(x)\,=\frac{3 \varepsilon \cc^2 }{x^2}+\,\frac{6 G_N M}{x^3}>0
\end{equation}
and, from \eqref{segnoW}, if $\Lambda>0$ then $\Phi({x}) >
0\;\forall x>0$.
\newline
Hence, in both cases, there are no barriers and we have a
monotonic expansion: \textit{if the $r$--shell is initially
expanding, it will go on expanding without limit, otherwise it
will collapse}.

\begin{figure}[h]
\centering
\includegraphics[width=0.87\textwidth]{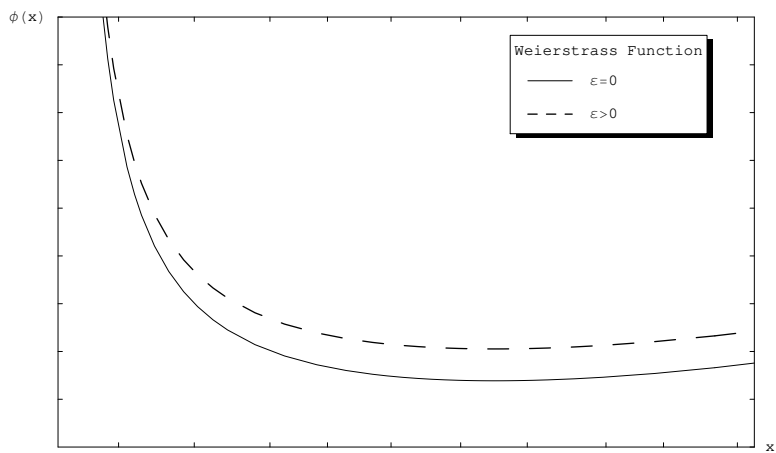}
\caption{The Weierstrass function $\Phi({x})$ for positive
$\Lambda$. Cases $\varepsilon=0$ and $\varepsilon>0$ are analyzed.
Here there are not inversion points: the motion $\forall\, x > 0$
is possible. }
\end{figure}

\subsubsection*{Positive $\Lambda$ and negative $\varepsilon$}
When $\varepsilon$ is negative, the function \eqref{W} becomes
\begin{equation}
W(x)\,=\,- \frac{3 |\varepsilon| \cc^2}{x^2}+\frac{6 G_N M}{x^3} \
,
\end{equation}
and it has a minimum at\,%
\foot{\ In fact its first $x$--derivative
$\quad
\frac{d W(x)}{d x}\,=\,\frac{6 |\varepsilon|
\cc^2}{x^3}-\frac{18G_N M}{x^4}
\quad$
is null when $x\,=\,x_M$. In addition, evaluating the second
derivative of $W(x)$ in $x_M$, we obtain the positive value
$\,\frac{2 |\varepsilon|^5\cc^{10}}{27 G_N^4 M^4}$. Hence $x_M$ is
a minimum point for the function $W(x)$.}
\begin{equation}\label{eq:softbarrier}
x\,=\,x_M\,=\,\frac{3 G_N M}{|\varepsilon|\cc^2} \ .
\end{equation}
Moreover, from Remark \ref{osservazione}, it is clear that when
$\Lambda$ assumes a suitable critical value $\Lambda_c$ defined as
\begin{equation}
\Lambda_c\,=\,-W(x_M)\,=\,\frac{|\varepsilon|^3 \cc^6}{9 G_N^2
M^2} \ ,
\end{equation}
the Weierstrass
function \eqref{eq:WeierstrassPhi} has a soft barrier\,%
%
in the point $x_M$, that we can also write as
\[
x_M = \left(\frac{3 G_{N}M}{\Lambda_c}\right)^\frac{1}{3}\ .
\]
%
\newline
Hence when $\Lambda$ assumes its critical value $\Lambda_c$, the
evolution of $r$--shells in LTB--models is static and stable.
\newline
Precisely the behavior of each $r$--shell depends by the initial
conditions: their evolution is really static if the initial
intrinsic radius equals $x_M$; on the other hand, the static
situation is a limiting situation: if $x_0\,\neq\,x_M$, the
r--shell go on expanding asymptotically approaching to the static
model with $x_M$ as intrinsic radius.
\par
If $\Lambda\,>\,\Lambda_c$, there are no barriers,
hence
\textit{if the $r$--shell is initially expanding $(\dot{x}_0>0)$,
it will go on expanding without limit}.
\par
Finally, when $0<\Lambda<\Lambda_c$, the Weierstrass function
\eqref{eq:WeierstrassPhi} admits two simple zeros
$$\bar{x}_{1}\,=\, 2 \sqrt{\frac{|\varepsilon|
\cc^2}{\Lambda}}\cos\frac{\alpha+4\pi}{3} \qquad \text{and} \qquad
\bar{x}_2\,=\, 2 \sqrt{\frac{|\varepsilon|
\cc^2}{\Lambda}}\cos\frac{\alpha}{3}\ ,$$
where $\alpha$ is defined by
$$\tan\alpha\,=\,\frac{\sqrt{-\Delta}}{-q}\ ,$$
with $q\,=\,\frac{3 G_N M}{\Lambda}$ and $\Delta\,=\,\frac{9G_N^2
M^2\Lambda-{|\varepsilon|}^3\cc^6}{\Lambda^3}$.
\footnote{Note that $\Delta$ is negative since
$0<\Lambda<\Lambda_c$. Moreover, since the tangent of $\alpha$ is
negative, $\alpha \in (\frac{\pi}{2}, \pi)$, hence
$\bar{x}_1<\bar{x}_2$.}
\newline
So when the initial intrinsic radius is less that $\bar{x}_1$ and
the $r$--shell is initially expanding, it will go on expanding
until the intrinsic radius reaches the maximal expansion point
$\bar{x}_1$; then it will contract back from $\bar{x}_1$ towards
the center of symmetry $O$ until it collapses in a finite time; on
the other hand, when the initial intrinsic radius is grater than
$\bar{x}_2$ and if the $r$--shell is initially expanding, it will
go on expanding.
\newline
Note that the open or closed evolution of the $r$--shell strongly
depends from the initial conditions.
\newline
Moreover, a particular situation is observed when the initial
intrinsic radius $x_{0}$ is such that $\bar{x}_1 < x_{0} <
\bar{x}_2$, since in this case the Weierstrass function is
negative: \textit{the initial intrinsic radius can't belong to the
interval $(\bar{x}_1, \bar{x}_2)$}.

\begin{figure}[h]
\centering
\includegraphics[width=0.87\textwidth]{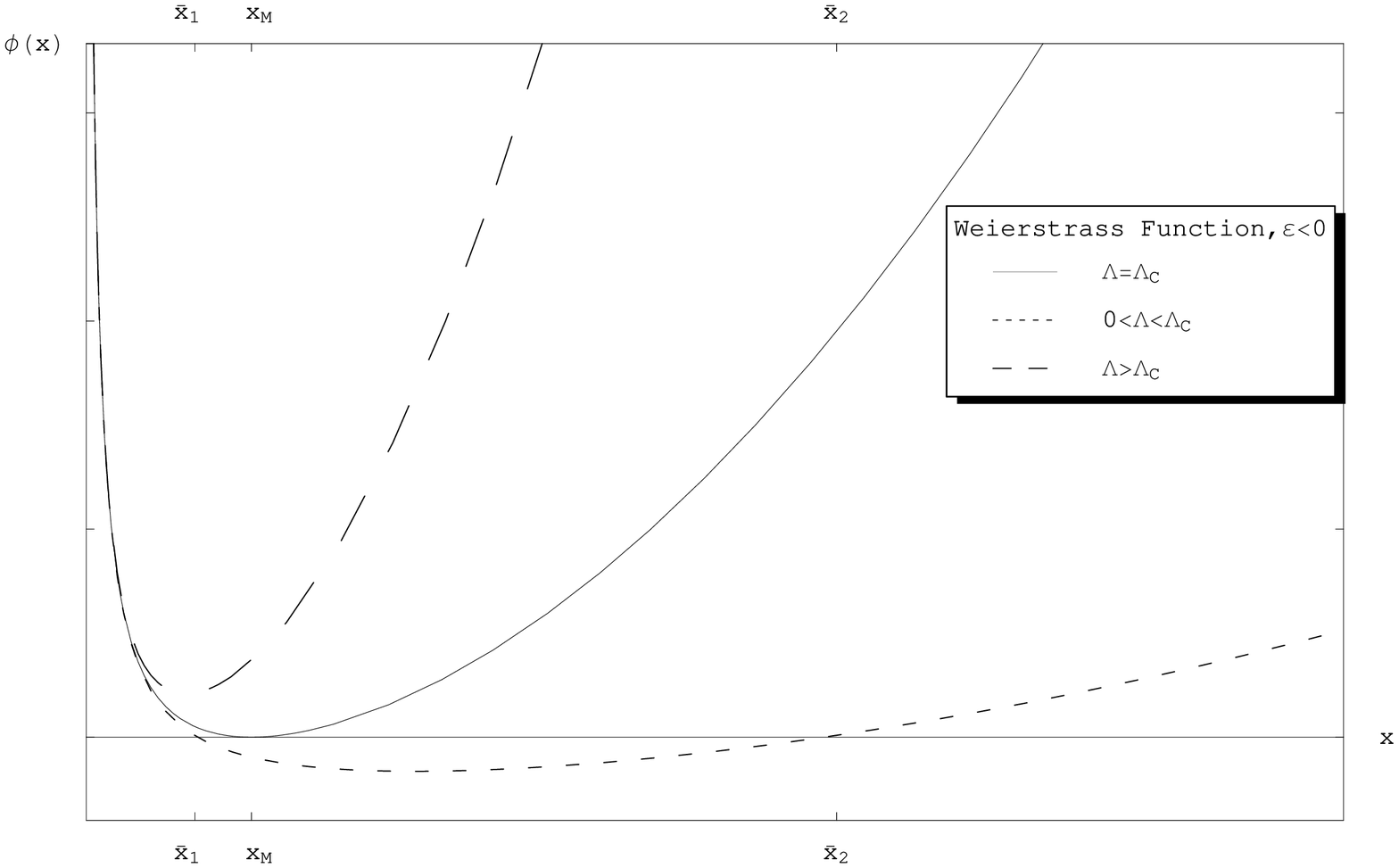}
\caption{ The Weierstrass function $\Phi({x})$ for positive
$\Lambda$ and negative $\varepsilon$. In this case the behavior of
the function depends on $\Lambda$. More precisely
\newline
i) when $\Lambda = \Lambda_c$, the function $\Phi({x})$ has a
multiple zero at ${x}_{_M}$. An equilibrium position or a limiting
position corresponds to the soft barrier. In the first case,
$x\,=\,{x}_{_M}$ is the unique possible position. In the second
case, the motion $\forall \,\, 0 < {x} < {x}_{_M}$ is possible;
\newline
ii) when $0<\Lambda < \Lambda_c$, the function $\Phi({x})$ has two
simple zeros at $\bar{x}_{1}$ and $\bar{x}_{2}$ (i.e. there are
two inversion points $\bar{x}_{1}$ and $\bar{x}_{2}$). The motion
is possible $\forall \,\, 0 < {x} \leq \bar{x}_{1}$ and $\forall
x\geq \bar{x}_{2}$;
\newline
iii) when $\Lambda > \Lambda_c$ the Weierstrass function has no
zeros. The motion $\forall\, x > 0$ is possible.}
\end{figure}

\begin{table}[ht] \label{table1}
\tbl{Evolution of of LTB $r$--shells for positive cosmological
constant and different signs of $\varepsilon(r)$.} {\scriptsize{
\begin{tabular}{c|lcc}
\hline & \quad Sign of $\varepsilon(r)$ & \qquad Evolution of
$r$--shells
 \\ \hline
$$ & $ $ \cr

 & {$\qquad \qquad \varepsilon=0$} & {\qquad Open }\\
\cr
\\
$ $ & $ $ \cr

$$ & $ \qquad
{\varepsilon<0}$ \quad $ {\Lambda\,=\,\Lambda_c}$ & $
{\text{Static solution as effective or limiting solution}}$\\
$$ & & $
{\text{In the second case the evolution is
open}}$\\
\cr
\\
{$\Lambda>0$} & $ \qquad {\varepsilon<0}$, \quad $
{\Lambda\,>\,\Lambda_c}$ & {Open}
\\
\cr
\\
$ $ & $ \qquad {\varepsilon<0}$\,, $ {0<\Lambda<\Lambda_c}$ &
{Closed if $x_0 \leq \bar{x}_1\,=\, 2 \sqrt{\frac{|\varepsilon|
\cc^2}{\Lambda}}\cos\frac{\alpha+4\pi}{3}$};\\
 & & {Open if $x_0 \geq \bar{x}_2\,=\, 2
\sqrt{\frac{|\varepsilon| \cc^2}{\Lambda}}\cos\frac{\alpha}{3}$}
\\
\cr
\\
$ $ & $ $ \cr
$$ & {$ \qquad \qquad \varepsilon>0$} &{Open}\\
\cr
$ $ & $ $ \cr $ $ & $ $ \cr \hline
\end{tabular}
} }
\end{table}

\begin{table}[ht] \label{table2}
\tbl{Evolution of LTB $r$--shells for negative cosmological
constant and different signs of $\varepsilon(r)$.} { \scriptsize{
\begin{tabular}{c|lcc}
\hline & \quad Sign of $\varepsilon(r)$ & \qquad Evolution of
$r$--shells
 \\ \hline
$$ & $ $ \cr

 & {$\qquad \qquad \varepsilon=0$} & {\qquad Closed}\\
\cr
\\
\\
{$\Lambda<0$} & $ {\qquad \qquad \varepsilon<0}$ & {\qquad Closed}
\\
\cr
\\
\\
$$ & {$ \qquad \qquad \varepsilon>0$} &
{\qquad Closed}\\
$ $ & $ $ \cr $ $ & $ $ & {(the maximum value of the intrinsic
radius depends also by $\Lambda$)} \cr
$ $ & $ $ \cr $ $ & $ $ \cr \hline
\end{tabular}
} }
\end{table}

\begin{figure}[h]
\centering
\includegraphics[width=0.87\textwidth]{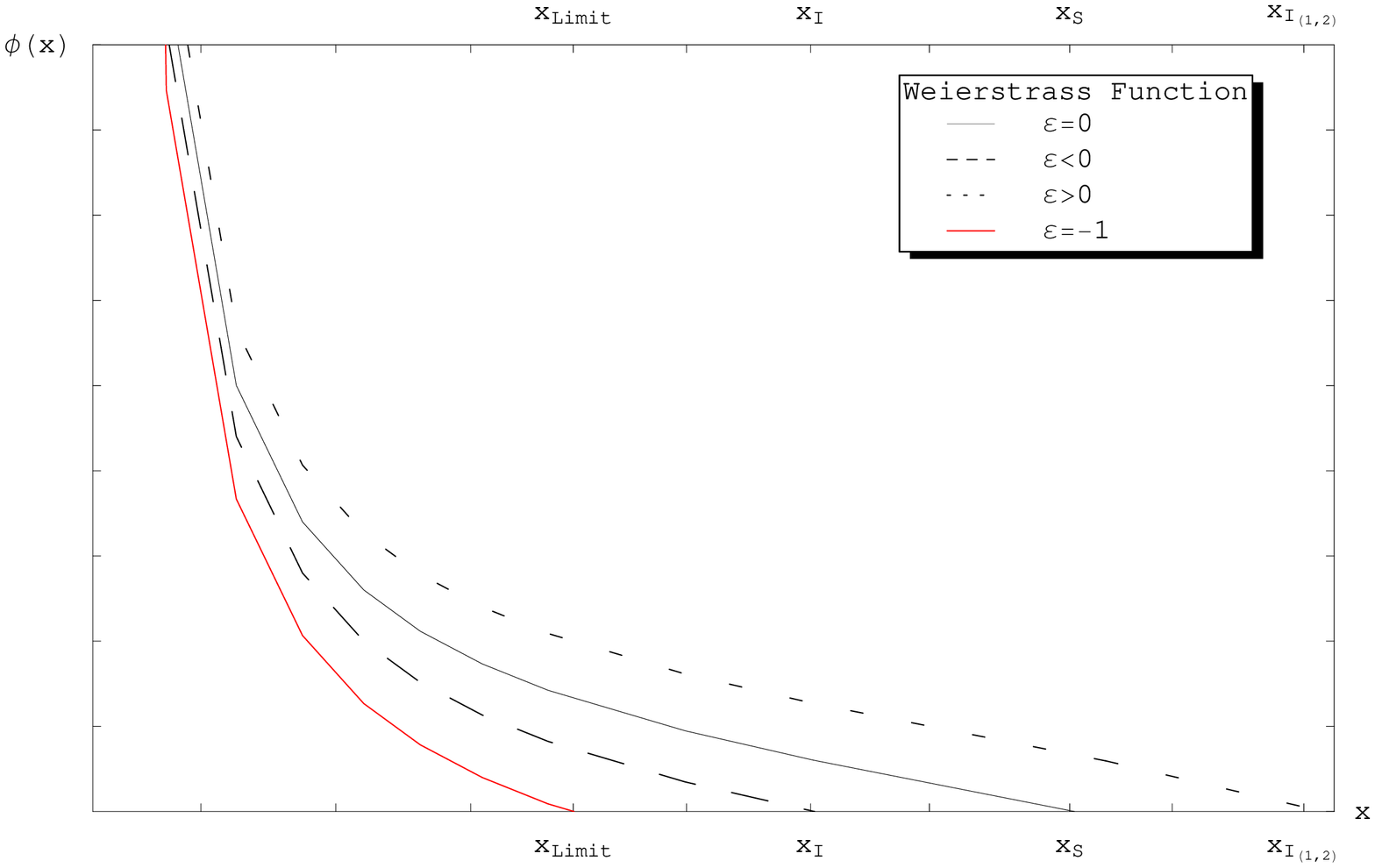}
\caption{ The Weierstrass function $\Phi({x})$ for negative
$\Lambda$ in three different cases. More precisely
\newline
i) when $\varepsilon = 0$, the function $\Phi({x})$ has a simple
zero at ${x}_{_S}$ (i.e. there is one inversion point ${x}_{_S}$)
and the motion $\forall \,\, 0 < {x} \leq {x}_{_S}$ is possible;
\newline
ii) when $\varepsilon < 0$, the function $\Phi({x})$ has a simple
zero at ${x}_{_I}$ (i.e. there is one inversion point ${x}_{_I}$)
and the motion $\forall \,\, 0 < {x} \leq {x}_{_I}$ is possible;
\newline
iii) when $\varepsilon > 0$ the Weierstrass function has a simple
zero. The value of this zero depends on $\Lambda$ as above
specified. It can be individuated by $x_{I_{1}}$ or by $x_{I_{2}}$
(i.e. there is one inversion point $x_{I_{1}}$ or $x_{I_{2}}$) and
the motion $\forall \,\, 0 < {x} \leq x_{I_{1}}$ (or clearly
$\forall \,\, 0 < {x} \leq x_{I_{2}}$) is possible;
\newline
finally, when $\varepsilon \leq -1$ it is represented the limiting
case $\varepsilon=-1$ when $x_{Limit}$ is the lower bound for the
set of inversion points $x_{_I}$. }
\end{figure}

\subsubsection{Negative cosmological constant}

Let us focus now on the qualitative behavior of $r$--shells for
negative value of $\Lambda$. In this case, the zeros of the
Weierstrass function may only be simple zeros, which correspond to
inversion points.
Even if the expression of the maximum value reached by the
intrinsic radius depends by the sign of $\varepsilon$, for each
analyzed situation (\textit{$\varepsilon=0$, $\varepsilon<0$} and
\textit{$\varepsilon>0$}) the model is closed.

\subsubsection*{Negative $\Lambda$ and null $\varepsilon$}
When $\varepsilon=0$ (Euclidean case), we find the barrier
\[
x_{_S} = \left(\frac{6 G_{N}M}{- \Lambda}\right)^{\frac{1}{3}} \ ,
\]
which is a simple zero of the Weierstrass function
\eqref{eq:WeierstrassPhi} and corresponds to an inversion point.
We have $\Phi(x)\leq 0$ for $x \geq x_{_S}$, hence the evolution
is possible for all values $0\leq x\leq x_{_S}$ .
\newline
\textit{When $\varepsilon\,=\,0$ and $\Lambda<0$, if the
$r$--shell is initially expanding, it will go on expanding until
the intrinsic radius reaches the invrsion point $x_{_S}$; then it
will contract back from $x_{_S}$ toward the centre of symmetry $O$
until it collapses in a finite time}.

\subsubsection*{Negative $\Lambda$ and negative $\varepsilon$}
When $\Lambda$ and $\varepsilon$ are both negative, there is a
barrier
\begin{equation}\label{xI}
\begin{split}
x_{_I} = \left(\frac{3G_{N}M}{|\Lambda|} +\sqrt{\frac{9 G_{N}^{2}
M^{2}}{|\Lambda|^{2}}+\frac{|\varepsilon|^{3}\cc^6}{|\Lambda|^{3}}}\right)^{\frac{1}{3}}
+
\\
+ \left(\frac{3G_{N}M}{|\Lambda|} -\sqrt{\frac{9 G_{N}^{2}
M^{2}}{|\Lambda|^{2}}+\frac{|\varepsilon|^{3}\cc^6}{|\Lambda|^{3}}}\right)^{\frac{1}{3}}
\ .
\end{split}
\end{equation}
The barrier $x_{_I}$ is a simple zero of the Weierstrass function,
that is an inversion point, so the model is again closed:
\textit{if the $r$--shell is initially expanding, it will continue
to expand until the intrinsic radius reaches its maximum value
$x_{_I}$; then it will contract back from $x_{_I}$ toward the
center of symmetry $O$ until it collapses in a finite time}.

\subsubsection*{Negative $\Lambda$ and positive $\varepsilon$}

Let's put, for sake of commodity,
\[
\Delta = \frac{9G_{N}^{2}M^{2}}{|\Lambda|^{2}} -
\frac{|\varepsilon|^{3}\cc^6}{|\Lambda|^{3}}\ .
\]
When
\[
|\Lambda| \geq \frac{\varepsilon^{3}\cc^6}{9{G_{N}}^{2}M^{2}}\ ,
\]
we have $\Delta\geq 0$ and there is a barrier
\[
x_{I_{1}} =
\left(\frac{3G_{N}M}{|\Lambda|}+\Delta^{\frac{1}{2}}\right)^{\frac{1}{3}}
+
\left(\frac{3G_{N}M}{|\Lambda|}-\Delta^{\frac{1}{2}}\right)^{\frac{1}{3}}\
,
\]
which is a simple
zero of the Weierstrass function, so corresponding to an inversion
point.
\par
Instead, when
\[
|\Lambda| < \frac{\varepsilon^{3} \cc^6}{9{G_{N}}^{2}M^{2}}\ ,
\]
$\Delta$ is strictly negative; let's put, for sake of commodity,
\[
\alpha =
\arctan\left(\frac{\sqrt{-\Delta}|\Lambda|}{3G_{N}M}\right)\ ,
\]
then there is a barrier
\[
x_{I_{2}} = 2\sqrt{\frac{3 \varepsilon
\cc^{2}}{|\Lambda|}}\cos\frac{\alpha}{3}\ ,
\]
which is again a simple
zero of the Weierstrass function, so corresponding to an inversion
point.

Hence, \textit{if the $r$--shell is initially expanding, it will
continue to expand until the intrinsic radius reaches the value
$x_{I_{1}}$ or $x_{I_{2}}$; then it will contract back from
$x_{I_{1}}$ or $x_{I_{2}}$ toward the center of symmetry $O$ until
it collapses in a finite time}.

\section{Concluding Remarks}
In this paper, we have analyzed a cosmological scenario based on
the spherically symmetric dust solutions of the Einstein
equations, that is LTB models with non--null cosmological
constant,
through an approach  strictly related to the Weierstrass method of
Classical Mechanics. In our case, it allows a systematic analysis
of LTB models according to the different signs and values of
$\varepsilon(r)$ and $\Lambda$.
%

In particular, while we can easily classify the LTB evolution with
null cosmological constant \cite{jim,grg}, namely the Universe is
open when $\varepsilon \geq 0$ and closed if $\varepsilon<0$, the
situation in presence of cosmological constant is much more
complex. First of all, one has to study the evolution for positive
and negative values of $\Lambda$ and, for each of these, relate
the evolution to the sign of $\varepsilon$. In particular, when
$\Lambda>0$, for $\varepsilon \geq 0$, the same global behavior of
LTB models with null cosmological constant is achieved: the
Universe is spatially open. In the other cases ($\varepsilon <0$)
the Universe can be open or closed and this feature strongly
depends on $\Lambda$ and on the initial conditions. The various
dynamical cases are summarized in Table I.

On the other hand, when $\Lambda<0$, for any value of
$\varepsilon$, LTB Universes present the same evolution: they are
always spatially closed. In particular, the maximum value reached
by the intrinsic radius depends only on $\varepsilon$ when
$\varepsilon \leq 0$ and also on $\Lambda$ when $\varepsilon>0$.
The dynamical behavior is more simple and summarized in Table II.


In addition note that, in the cases corresponding to spatially
closed Universes, we are not only able to select such a
characterizing feature, but also to obtain the exact value of the
maximum that the $r$--shell can reach.

As concluding remark, it is worth noticing that the method
outlined here can be, in principle, applied any time that
cosmological dynamical system can be recast along the Weierstrass
function and an effective cosmological constant can be defined.
For example, it could be possible to apply this approach to
cosmological models including scalar fields \cite{capozziello}.

\end{document}